\documentclass[letterpaper]{raa}

\usepackage{amssymb}
\usepackage{multirow}

\usepackage{graphicx}
\usepackage{enumerate}
\usepackage{amsmath}

\usepackage{graphicx,times,amssymb,hyperref}             
\input{epsf.sty}                        
\input{psfig.sty}                       
\usepackage[totalwidth=480pt,totalheight=680pt]{geometry}

\usepackage{enumerate}

\def\erg{\,{\rm erg}}

\def\cm{\,{\rm cm}}

\def\ergscm2 {erg\,s$^{-1}$cm$^{-2}$}

\def\cm2 {cm$^{-2}$}

\def\aap {A\&A}
\def\apj {ApJ}

\begin{document}

\title{Quark-Novae Ia in the Hubble diagram: Implications For Dark Energy}

 \volnopage{Vol.0 (200x) No.0, 000--000}      
   \setcounter{page}{1}          

   \author{Rachid Ouyed    \inst{1} 
   \and Nico Koning  \inst{1}
   \and Denis Leahy  \inst{1}
   \and Jan E. Staff  \inst{2}
   \and Daniel T. Cassidy  \inst{3}
      }

   \institute{Department of Physics and Astronomy, University of Calgary, 
2500 University Drive NW, Calgary, Alberta, T2N 1N4 Canada; {\it rouyed@ucalgary.ca}\\
        \and
Department of Physics and Astronomy, Macquarie University NSW 2109, Australia
       \and
Department of Engineering Physics, McMaster University, Hamilton, Ontario, Canada L8S 4L7
   }

   \date{Received~~2013 month day; accepted~~2013~~month day}

\abstract{The accelerated expansion of the Universe was  proposed through the use of Type-Ia SNe as standard candles.  The standardization depends on an empirical correlation between the stretch/color and peak luminosity of the light curves.  The use of Type Ia SN as standard candles rests on the assumption that their properties (and this correlation) do not vary with red-shift. We consider the possibility that the majority of Type-Ia SNe are in fact caused by a Quark-Nova detonation in a tight neutron-star-CO-white-dwarf binary system; a Quark-Nova Ia.  The spin-down energy injected by the Quark Nova remnant (the quark star) contributes to the post-peak light curve and neatly explains the observed correlation between peak luminosity and light curve shape.
We demonstrate that the parameters describing Quark-Novae Ia are NOT constant in red-shift.  Simulated Quark-Nova Ia light curves provide a test of the stretch/color correlation by comparing the true distance modulus with that determined using SN light curve fitters.   We determine a correction between the true and fitted distance moduli which when applied to Type-Ia SNe in the Hubble diagram recovers the $\Omega_{\rm M} = 1$ cosmology.  We conclude that Type-Ia SNe observations do not necessitate the need for an accelerating expansion of the Universe (if the observed SNe-Ia are dominated by QNe-Ia) and by association the need for Dark Energy.
}

   \authorrunning{Ouyed  et al.}            
   \titlerunning{Quark-Novae Ia in the Hubble diagram: Implications For Dark Energy}  

   \maketitle

\section{Introduction}

A surprising finding in the past decade is  the (late-time) acceleration of the expansion of the Universe using Type Ia SNe (hereafter SNe-Ia) (Riess et al., 1998; Perlmutter et al., 1999).   The acceleration seems to have kicked in at a redshift of about $\sim$ 0.5 due to a mysterious ``force" termed Dark Energy (DE) (accounting for more than 70\% of the overall density of our universe). This has led to a paradigm shift, making astrophysicists rethink the simple picture of a decelerating universe filled with matter. Nature thus once more seems to surprise us by hinting at this puzzling component acting as a dominant negative pressure tearing our universe apart.   Although the energy density of DE is very low ($< 10^{-29}$ g cm$^{-3}$) it is uniformly distributed  (i.e. a non-clustering form of energy) which dominates on large scales.

Quantum vacuum energy was postulated as one possible candidate for DE, but linking DE with Einstein's cosmological constant (Einstein 1917) imposes extreme fine tuning of the physics at the Planck time and  observations indicating a vacuum energy that is at least 60 orders of magnitude smaller than we think it should be, given our knowledge of how particles interact (e.g. Weinberg 1989; Carroll 2001; Frieman et al. 2008). 
One might speculate that General Relativity (GR) breaks down on cosmological scales; if so then GR needs to be replaced with a more complete theory (Milgrom 1983;  Bekenstein 2004; Harko \& Cheng 2005;
see also Sotiriou \& Faraoni 2010 for  a review).  Other theories and models involve a slowly rolling scalar field generating a negative pressure (e.g.
 Ratra \& Peebles 1988; Turner \& White 1997; Chiba et al. 2009 to cite only a few). Any of these ``solutions" would fundamentally change our grasp of the laws governing the physical evolution of our universe. Understanding the cosmic acceleration has, therefore, taken center stage in modern science with deep connections to both astrophysics and particle physics.

This  discovery was first arrived at by assuming SNe-Ia are ``standardizable" candles that are driven purely by $^{56}$Ni.  SNe-Ia result from the thermonuclear explosion of a carbon-oxygen (CO) white dwarf (WD) (Hoyle \& Fowler
1960).  Explosion models of carbon-oxygen CO  WD  give spectral features
 which successfully fit  spectra  of  SNe-Ia  (Nomoto 1982a).  In the traditional picture, accretion from a companion (a main sequence star, giant star or another WD;  e.g. Whelan \& Iben 1973; Webbink 1984; Iben \& Tutukov 1984) drives the WD toward the Chandrasekhar limit (Chandrasekhar 1934) triggering the explosion.  
  Models involving the (off-center detonation) explosion of a sub-Chandrasekhar WD  have also been considered (e.g. Woosley et al. 1980;
  Nomoto 1982b; Livne \& Arnett 1995; Fink et al. 2010).  Once the explosion has occurred, the physics governing the expansion of the ejecta is well understood; the expansion velocity is linked to the burning of CO to $^{56}$Ni while the light curve is fed by radioactive decay of $^{56}$Ni and later $^{56}$Co. The energetic gamma-rays from the radioactive decay  heats up the ejecta, and are then re-emitted in the optical, with a direct connection between the light curve peak (which occurs at about 18 days after the explosion) and the amount of $^{56}$Ni processed (Arnett 1969\&1982; Woosley \& Weaver 1986;  Khokhlov 1989\&1991).

Observations of SNe-Ia spectra and light curves (LCs) allow one to determine the redshift of the exploding WD and its intrinsic peak luminosity; the LC is ``indirectly" used to infer the corresponding  luminosity distance $d_{\rm L}(z)$. A correlation between the peak absolute magnitude and the shape of the light curve was first noted by Phillips (1993).  This ``Phillips relationship" along with correlations in B-V colour (Riess et al. 1996; Tripp 1997) have been used to reduce the scatter in the intrinsic luminosity (from about 1 mag to about 0.1 mag; Kessler et al. 2013).  These ``standardizable" candles as they are referred to in the literature are at the heart of this revolution.  With the appropriate correlations taken into account, the Hubble diagram constructed from the SNe-Ia shows a clear deviation from a flat, matter-filled universe; i.e. $d_{\rm L}(z)$ is dominated by an accelerating component.

The nature of the progenitors and the physics behind the detonation of SNe-Ia are still not well understood.  Furthermore,  among the population of SNe-Ia exist outliers which in some cases call for care and caution when  interpreting their observed spectra and light curves (e.g. SN 1991T (Filippenko et al. 1992
and Phillips et al. 1992) ;  SN 2006bt (Foley et al. 2010);  see \S\ 4.3 in Ouyed \& Staff (2013) for a discussion and other peculiar SNe-Ia).  Nevertheless the use of these ``adjustable beacons" as cosmological probes seems to have survived the test of time and their roles as distance indicators and DE messengers is becoming a reality. Furthermore, as we discuss below, other probes (complementary observations) in conjunction with and/or independently from SNe-Ia seem to also hint at the existence of DE.

\subsection{The concordance model}

Modern observational cosmology has converged in the last two decades on what is known as the ``concordance model'' (Ostriker \& Steinhardt 1995) -- a spatially-flat $\Lambda$CDM parameterization where deviations from a linear Hubble relation derived from SNe-Ia, the present-day observed distribution of baryons combined with mass-to light ratios, and the angular power spectrum of fluctuations in the Cosmic Microwave Background (CMB) temperature are all consistent in a narrow range of $\Omega_{\rm M}$ and $\Omega_{\Lambda}$ for a Hubble constant which is tightly-constrained in the near-field.

A great deal of work during the intervening years to improve the observational tests and constraints of the ``concordance model'' has culminated in the cosmological parameters paper by the Planck team (Ade et al. 2013). They find that the standard six-parameter $\Lambda$CDM cosmology is sufficient to explain all new observational constraints based on CMB temperature power spectra and lensing-potential power spectra. 

Of the three constraints in the ``concordance model'', two are different linear combinations of $\Omega_M$ and $\Omega_{\Lambda}$. SNe-Ia constrain the Hubble constant in the near-field but are sensitive to the rate of change of the Hubble constant - which can be produced by a non-zero $\Lambda$ -
at redshifts of z=1 and greater (Riess et al. 2007). The angular power spectrum of temperature fluctuations in the CMB indicates that the curvature parameter is zero and consequently, its contribution to the total $\Omega$ must be zero. An $\Omega$ = 1 requires that $\Omega_{\rm M}$ $+$ $\Omega_{\Lambda}$ = 1. The distribution of matter, both baryonic and non-baryonic, is found to contribute $\Omega_{\rm M}$ $=$ 0.315,
and consequently, $\Omega_{\Lambda}$ is the largest contribution to the energy density of the Universe at a level marginally below 70\%.

The constraining power of SNe-Ia on the ``concordance model'' relies on the assumption that such supernovae do not have properties that evolve with time (lack of ``evolution'') and that, if other scenarios exist to produce transients of similar luminosity, those transients are few in number relative to normal Type Ia events, or are easily distinguished from them using existing survey strategies. Evolution of the class of SNe-Ia as a source of systematic error appears to be well-constrained observationally and therefore unlikely (Blondin et al. 2006).

\subsection{What if ... ?}

Given the lack of understanding into the nature of DE and its discordance with our perception of how the Universe works and evolves, we owe it to ourselves as a community and as scientists to explore every possible avenue that could lead to SNe-Ia-like explosions (i.e. involving the explosion of a CO WD;
hereafter COWD) that follow a Phillips-like relationship but might not be  ``standardizable" (i.e. cannot be used as standard candles). If such events exist in nature, and if they are among standard SNe-Ia, one should be concerned about their effect on the Hubble diagram. It was shown by Ouyed \& Staff (2013) that a Quark-Nova (QN) occurring in a tight neutron star (NS) COWD (hereafter NS-COWD) binary may lead to such an explosion.  However, in the Ouyed \& Staff (2013) scenario, there is an additional energy source which  is provided by the spin-down power\footnote{Energy injection  by spinning-down quark stars  was previously  
 suggested  in the context of Gamma-Ray Burst  afterglows (Dai \& Lu 1998). Furthermore,
 we note that the conversion of a NS to a strange star by accretion 
 in a low-mass X-ray binary was also investigated in the context of Gamma-Ray Bursts (Cheng \& Dai 1996).} of the quark star (the QN remnant). It is the aim of this paper, then, to investigate the extent to which this scenario affects the  Phillips relationship and thus the 
  Hubble diagram. As it turns out the spin-down energy  can ``fool" the light curve fitters used in calibration, and can shift 
   the true moduli in the Hubble diagram. We show that appropriate corrections   put the observed SNe-Ia back
   onto the $\Omega_{\rm M}=1$ universe thus taking away the need for DE.  

The paper is organized as follows: In \S\ 2 we briefly introduce the QN and describes its key features.
In \S\ 3 we describe the QN-Ia which results from the explosion of a COWD in 
 in a tight NS-COWD binary.  An estimate of the QN-Ia rate is given with emphasis
 on the Common Envelope channel as the path to QN-Ia systems. The resulting QN-Ia 
 LC  is described in  \S\ 4 and is compared to pure $^{56}$Ni powered SN-Ia lightcuves.
 The dependence of QNe-Ia LC with redshift is outlined in \S\ 5. The implications to the Hubble
 diagram are highlighted in \S\ 6. A discussion of the impact of QNe-Ia  on Cosmology and
  DE is given in \S\ 7  where we also list our model's limitations. We conclude in \S\ 8.
  
\section{Quark-Nova}
\label{sec:theQN}

A QN is an explosion resulting from the conversion of a NS to a quark star (QS) (Ouyed et al. 2002; Ker\"anen et al. 2005).  The extreme conditions within the core of a NS may result in a phase transition from hadronic matter to the theoretically more stable up-down-strange (UDS) matter (Itoh 1970, Bodmer 1971, Witten 1984; see  also Terazawa 1979).  The critical mass of the NS, $M_{\rm NS, c}$, needed to reach the quark deconfinement density in the core
and cause a QN explosion varies from $\sim 1.6M_{\odot}$  up to $\sim 1.9M_{\odot}$ 
and higher for a variety of equation of states (EoSs); see Staff et al. (2006).  
 A   typical gravitational mass of a NS at birth is $M_{\rm NS, b} \sim  1.4M_{\odot}$ where the subscript ``b" stands for birth.
 Therefore, the amount of  baryonic mass 
needed to drive a  NS  above  $M_{\rm NS, c}$ and experience a QN episode is
 $\Delta M_{\rm QN}= M_{\rm NS, c}/(1-2 G M_{\rm NS,c}/R_{\rm NS,c} c^2)^{1/2}  - M_{\rm NS, b}/(1-2 G M_{\rm NS,b}/R_{\rm NS,b}c^2)^{1/2} \sim 0.3$-$0.7M_{\odot}$; $G$ and $c$ are the gravitational constant and the speed of light, respectively.

The critical NS  mass limit could be made
higher   (to account for the recent $\sim 2M_{\odot}$ observed by Demorest et al. 2010) by setting the quark deconfinement density above five times the nuclear saturation density or  using a different/stiffer equation of state of neutron matter (see Staff et al. 2006 and \S\ 2.1 in Ouyed \& Staff 2013 for a discussion). 
 A high $M_{\rm NS, c}$ in our model pushes the amount of material to be accreted by the NS  to extreme values   (see the discussion in \S\ \ref{sec:limitations}) but does not change our findings and the conclusion of the paper. 
 The QN remnant is a quark star with mass\footnote{Massive ($\sim 2M_{\odot}$ or heavier)  quark stars can exist when one considers an interacting phase of quarks  (Alford et al. 2007).} close to the parent NS since the QN ejects at most $10^{-2}M_{\odot}$ of neutron-rich and iron-rich material.

In a QN, besides the neutrino emission, a  photon fireball is generated since the temperature of the quark core is large enough at the time of formation (much above the quark plasma frequency) to sustain large photon emissivities (Vogt et al. 2004; Ouyed et al. 2005). The photon energy deposited in the outer layers of the NS (including the crust), will impart a large momentum  leading to strong and ultra-relativistic mass ejection (Ouyed \& Leahy 2009).
 The energy released through this conversion, coupled with the ensuing gravitational collapse of the NS ejects on average $10^{-3} M_{\odot}$ of material at kinetic energies $> 10^{52}$ erg (see Ouyed et al. 2013 for a review).   This QN ejecta consists of neutron-rich and iron-rich material representative of the NS outermost layers.  The fate of this relativistic, very dense,  ejecta (with an average Lorentz factor $\Gamma_{\rm QN}\sim 10$) leads to a variety of observables

   If the QN occurs within weeks of the original SN, the QN ejecta can interact with the SN remnant, leading to a super-luminous SN (Leahy \& Ouyed 2008, Ouyed et al. 2009, Ouyed et al 2012) or a double-humped SN (Ouyed et al. 2013).  In fact the double-humped SN was predicted by the QN model three years before it was observed as SN 2009ip and SN 2010mc.  Most recently, the consequences of a QN as part of a WD binary system was considered by Ouyed \& Staff (2013); an event resulting in the explosion of the WD (the QN-Ia; see below).  Other astrophysical implications and signatures of QNe in binaries have been investigated in Ouyed et al. (2011a\&b).
  If the QN ejecta has insufficient energy, it may fallback and form a shell or Keplerian ring around the QS resulting in what we observe as soft-gamma repeaters (SGRs), or anomalous x-ray pulsars (AXPs) (see e.g. Ouyed et al. 2007a\&b).   We refer the interested reader to \S\ 2 in Ouyed \& Staff (2013) for more details on the physics and  observational signatures of QNe. 

\section{QN-Ia}
\label{section:qnia}

Ouyed \& Staff (2013) considered the scenario in which a NS and COWD form a tight binary system, where mass transfer from the WD to the NS would occur.  The accreted mass would drive the NS above the critical mass of $M_{\rm NS, c}$ and undergo a QN explosion.  
The QN ejecta would then collide with the WD, driving a shock that triggers carbon burning under degenerate conditions leading to the thermo-nuclear explosion of the WD;  extreme shock compression allows runaway burning even for  low-mass (i.e. $< 0.5M_{\odot}$) WD in the QN-Ia model.   The conditions of the ejecta as it hits the WD have been presented in \S\ 2.3 in Ouyed \& Staff (2013).
 The resulting SN (the Quark-Nova Ia; hereafter QN-Ia) has features (spectra and light curves) that are very much like standard Chandrasekhar-mass,
and sub-Chandrasekhar mass,  thermo-nuclear  explosions with one very important difference: an extra energy source from the spin-down power of the QN compact remnant (i.e. the QS); see  Ouyed \& Staff (2013) for more details.

\subsection{The common envelope channel}
\label{sec:CE}

In our model, a QN-Ia is the result of a common-envelope (CE)  evolution  of a NS-AGB pair. 
The progenitor stellar system of this binary is one containing a massive star ($M>8M_{\odot}$) and an intermediate mass star ($1.0M_{\odot}<M_2 <8M_{\odot}$). The massive star rapidly evolves\footnote{In $<\sim 20$Myr, thus we can safely ignore this small time in the rate calculation.} and explodes to produce a NS whereas the intermediate mass star evolves much more slowly into a giant star. Before the intermediate mass star fully evolves, as it would if it were a single star, it overflows its Roche lobe (if it is in a sufficiently close binary), leading to a CE phase and shrinking of the binary orbit  (Paczynski 1976; van den Heuvel 1976; Taam et al.  2000; Tauris et al. 2000).

In a fraction of cases, the binary may successfully eject the AGB envelope and emerge as a short period ($\sim$ a few hours) NS-COWD
system (e.g.  Tauris et al. 2000 and references therein).  Using the energy formalism it has been argued that the NS accretes a fraction of the envelope
mass leading to ejection (see \S\ 4 of Brown 1995; see also Armitage \& Livio 2000). Most of the mass is accreted at  hyper-accretion rates late during the spiral-in  on a timescale of order a year which corresponds to the dynamical timescale of the AGB envelope.   There are many uncertainties on
 the details of the in-spiral phase and envelope ejection  (e.g. Izzard et al. 2012; Ivanova et al. 2013) but general arguments can be used 
 that gives the mass accreted onto the NS ($\Delta M_{\rm  NS, CE}$) in this phase to scale as the envelope mass; i.e. $\Delta M_{\rm NS, CE}\propto M_{\rm env.}\propto  M_2$.   {\it This phase is important for the QN-Ia because a higher mass companion $M_2$ results in a more massive NS which then
  requires less subsequent accretion from the WD to trigger the QN.}
    We write   
   \begin{equation}
   \Delta M_{\rm NS, CE}= \psi_{\rm CE} M_2 \ ,
   \end{equation}
   with $\psi_{\rm CE}=0.2$ in this work  (see \S\ \ref{sec:offset}).

In cases where the NS accretes  enough mass to undergo a QN explosion (or go directly to a black hole) while it is still in the CE phase,
 the explosion would be buried  and would not result in a SN-Ia-like explosion (instead this could result in a typeII-like explosion).  Here we are only interested in cases where the CE
 is ejected leaving a tight NS-COWD binary.  This means we require $\Delta M_{\rm NS,CE} < \Delta M_{\rm QN}$ with $\Delta M_{\rm QN}$ defined in \S\ \ref{sec:theQN}. 
 
The orbital period after   the CE ejection is on the order of a few hours (e.g. Tauris et al. 2000).
  The binary orbit needs to shrink further for the WD to overflow its Roche-lobe and accrete onto
 the now more massive NS which triggers the   QN-Ia.  This further shrinkage of the orbit is driven by
 gravitational radiation losses and occurs on timescales of  $\tau_{\rm GW} \sim 10^8$ years (e.g. Landau \& Lifshitz 1975; see also
 \S\ 5 in Ouyed \& Staff 2013).    
 
 It is important to note that the observed NS-COWD binaries are those that emerged from the CE
 with long enough periods that the gravitational radiation does not cause them to merge
 in a less than 1 Gyr.    In other words, the observed NS-COWDs systems may have never entered the CE phase
  or if they did, they would have  emerged from
the CE with sufficiently long periods that they have not merged yet.

\subsection{Rates}
\label{sec:rates}

Over the past 20 years, the cosmic history of star formation has been measured back to as early as 500 Myr after the Big Bang (e.g. Dunlop, 2011). At early times the star formation rate was much higher than at present, rising from 500 Myr then peaking near an age of 2-4 Gyr (at a redshift of $\sim$2-3), then steadily declining to the present day. The star formation rate at  $2  <  z < 3$ was a factor 20-30 higher than the present rate.

The fraction of massive star progenitors in binary systems with companions in the intermediate mass (IM) range of $1.0M_{\odot}<M<8M_{\odot}$ is $f_{\rm IM}\sim$20\% \footnote{Most massive stars form in binary systems and a Salpeter mass function gives $\sim$20\% of the companions will be in the $1.0M_{\odot}<M<8M_{\odot}$ range, most will be at the lower end.}. We take an estimate of those binaries that survive the SN explosion as $f_{\rm survival}\sim20$\% (e.g. 
Kalogera 1996).   The companion in the surviving NS-intermediate mass star binaries expands rapidly when it enters its red giant (RG) then its  AGB phase.
  We take  the fraction of systems  $f_{\rm AGB}\sim$ 20\%  which enter the CE phase during the AGB phase (Mathieu 1994; see also Duquennoy \& Mayor  1991).
   Those that enter the CE during the RG phase will lead to a NS-HeWD system which do not result in a QN-Ia (see however \S\ \ref{sec:predictions}).
    Finally,  to end up as a QN-Ia the binary needs to survive the CE spiral-in phase and emerge as a short
    period  NS-COWD binary.  The fraction or AGB-CE systems  leading to a QN-Ia binary we take to be $f_{\rm CE, QNIa}\sim$ 10\% (see \S\ \ref{sec:limitations}).
    The net fraction of SN II explosions that end up as QN-Ia progenitor binaries is then
    
\begin{eqnarray}
\label{eq:fQN-rate}
f_{\rm SN, QNIa}&&\sim   f_{\rm binary}\times f_{\rm survival}\times f_{\rm IM}\times f_{\rm AGB}\times f_{\rm CE, QNIa}\notag \\
 &\sim&  0.5\times 0.2\times 0.2\times 0.2\times 0.1 \sim  4\times 10^{-4}\ . 
\end{eqnarray}

To estimate the QN-Ia rate, we take the current rate of NS formation in type II SN for our Galaxy of $r_{\rm SNII,0}\simeq$ 1 per 60 years (with about half of them in binaries; $f_{\rm binary}\sim$ 50\%). We approximate that the majority of progenitor binary systems were formed at $z_{\rm burst}\sim 2.5$ over a period of $\sim$ 2 Gyr. Since the star formation rate has dropped significantly since $z_{\rm burst}\sim 2.5$, most progenitor systems will have been formed during that $\tau_{\rm burst}\sim$ 2 Gyr period. For a typical Milky Way type galaxy, the number of QN-Ia progenitor systems formed will be:

\begin{equation}
N_{\rm prog}=f_{\rm SN,QNIa}\times \left( \tau_{\rm burst} \times f_{\rm SFR(z=2.5)} \times r_{\rm SNII,0} \right) \ ,
\end{equation}

\noindent where $f_{\rm SFR(z=2.5)}\simeq$20-30 is the increase in star formation rate at $z=2.5$ over current $z=0$ rates, and $f_{\rm SN,QNIa}\sim  4\times 10^{-4}$, estimated in the previous paragraph.  
With the fiducial values, the estimated number of progenitors in a galaxy is $N_{\rm prog}\sim 4\times 10^{5}$.

These NS-intermediate mass progenitor binaries evolve into the CE phase and thus into QNe-Ia over time. The main sequence mass vs. WD core mass relation that we use is (e.g. Weidemann,  2000; Kalirai et al. 2008; Catal\'an et al. 2008 to cite only a few):

\begin{equation}
\frac{M_{\rm WD}}{M_{\odot}} \sim 0.4+0.1\times \frac{M_{\rm MS}}{M_{\odot}}\ .
\label{eq:MWD_MS}
\end{equation}

Those systems with more massive companions to the NS evolve more quickly so that the QNe-Ia with more massive WD are produced earlier (higher red-shift) than those at the current time. Main sequence lifetimes are well-known and a useful analytic approximation is a power law (e.g. see Prialnik, 2000): 

\begin{equation}
\tau_{\rm MS}=10~\rm{Gyr} \times (M/M_{\odot})^{-\alpha}\ .
\label{eq:tms}
\end{equation}

For a mass range of $1.0M_{\odot}<M<8M_{\odot}$, the exponent $\alpha$ is between $3.0$ and $4.0$. We use a middle value of -3.5 in this work. This main sequence lifetime is by far the largest component of the time delay to form QN-Ia.  We now proceed to calculate the rate of QNe-Ia vs. time since the starburst.  

Using the Salpeter mass function (Salpeter 1955), the number of QN-Ia progenitor systems is:

\begin{equation}
N_{\rm prog}=\int_{1}^{8}\frac{A}{M_\odot} \left(\frac{M}{M_{\odot}}\right)^{-2.3} dM\ ,
\label{eq:Nprog}
\end{equation}
which gives $A\sim 5.2\times 10^5$.

The QN-Ia rate can be expressed as:

\begin{equation}
\frac{dN_{\rm QN-Ia}}{dt}= \frac{-dN_{\rm prog}}{dM} \frac{dM}{dt} \ .
\label{eq:dNdt}
\end{equation}

Making use of Equation \ref{eq:tms} we have:

\begin{equation}
\frac{dt}{dM} = \frac{d\tau_{MS}}{dM} = -\frac{10~{\rm Gyr}~\alpha}{M_\odot} \left(\frac{M}{M_\odot}\right)^{-\alpha-1} \ .
\label{eq:dtdM}
\end{equation}

Substituting Equations \ref{eq:dtdM} and \ref{eq:Nprog} into \ref{eq:dNdt} and using \ref{eq:tms} gives:

\begin{equation}
\frac{dN_{\rm QN-Ia}}{dt} = \frac{A}{10~{\rm Gyr}~ \alpha} \left(\frac{t}{10~\rm{Gyr}}\right)^{-1+1.3/\alpha} \ .
\end{equation}

The reason for the weak time-dependence of the QN-Ia rate (i.e. the $(-1+1.3/\alpha)$ power law above) is a competition between two factors: (i) the Salpeter mass-function which favours low-mass stars thus increasing the rate of large main-sequence lifetimes or large time-delays (i.e. low red-shift); (ii) countering this the steeply shorter lifetime for massive progenitors which increases the rate at low time delays (i.e. higher red-shift).  The former has a $\sim 2.3$ power law while the latter has a $\sim 3.5$ power law dependence on mass. 

For $\alpha=3.5$ and $A\sim 5.2 \times10^5$, we obtain $dN_{\rm QN Ia}/dt \simeq  1.5\times 10^{-5} (t/\rm{10Gyr})^{-0.63} \rm{yr}^{-1}$. In the Einstein- de-Sitter (matter-dominated flat universe) age can be related to red-shift as (Einstein \& de Sitter 1932):

\begin{equation}
\tau(z) = \frac{2}{3H_0(1+z)^{3/2}} = 9.3~{\rm Gyr}~(1+z)^{-3/2} \ ,
\label{eq:tz}
\end{equation}

\noindent where we have used $H_0 = 70.4\ {\rm km~s^{-1} Mpc^{-1}}$ (Komatsu et al. 2011).  The time elapsed since the bursting era, $z_{\rm burst}$ is therefore:

\begin{equation}
t(z)= 9.3~\rm{Gyr}~\times \left[\frac{1}{(1+z)^{3/2}}-\frac{1}{(1+z_{\rm burst})^{3/2}}\right] \ .
\end{equation}

The QN-Ia rate in today's universe ($z=0$) is therefore $\sim 2\times 10^{-5} ~\rm{yr}^{-1}$ per galaxy when taking
$z_{\rm burst} \sim 2.5$.  {\it  This rate is about a factor 50 smaller  than the observed rate of SNe-Ia (e.g. Dilday et al. 2008; Maoz \& Mannucci 2012; Quimby et al. 2012).  However, we note that   $f_{\rm AGB}$,  $f_{\rm IM}$,  $f_{\rm survival}$
 and $f_{\rm CE, QNIa}$ could reasonably be each a factor of $\sim 2$ higher,
  bringing the QN-Ia rate within  the observed SNe-Ia rate (see discussion in \S\ \ref{sec:limitations})}.


\section{Lightcurves}
\label{section:lightcuves}

The QN-Ia light curve will be powered by a combination of $^{56}$Ni decay and energy deposited by the spin-down of the QS.  The mass of the WD (M$_{\rm WD}$), the fraction of $^{56}$Ni produced (f$_{\rm Ni}$) and the spin-period of the QS (P$_{\rm QS}$) will all effect the peak and shape of the resulting light-curve.

We model the pure $^{56}$Ni light curve (L$_{\rm Ni}$) using the semi-analytical expressions given by Equations 9 and 10 in Chatzopoulos et al. (2012) (we include the $\gamma$-ray leakage term with Thomson opacity).  For the spin-down contribution (L$_{\rm SD}$), we use Equation 13 from Chatzopoulos et al. (2012).  The bolometric luminosity of our QN-Ia is therefore $L_{\rm QN-Ia} = L_{\rm Ni} + L_{\rm SD}$.
This bolometric luminosity yields a photospheric temperature assuming a photospheric radius given by Arnett (1982) where the photosphere is approximated by $R-(2/3)\lambda$ with $\lambda$ being the mean-free-path and $R=v_{\rm exp}t$.  The temperature gives a black-body spectrum which we use to construct B and V band QN-Ia light curves.   In the left panels of Figure \ref{fig:phillips_1} we simulate QN-Ia light curves using the prescription above for several different QS spin-periods with $M_{\rm Ni} = 0.27 M_\odot$ (top panels) and $M_{\rm Ni} = 0.22 M_\odot$ (bottom panels).  

\subsection{Phillips Relation}

An empirical correlation between the shape and peak absolute magnitude of SNe-Ia light curves was noted by Phillips (1993).  In particular, the change in B apparent magnitude from time of maximum light to 15 days later ($\Delta m_{\rm B,15}$) has been shown to correlate linearly to the peak absolute magnitude ($M_{\rm B,max}$) (Phillips 1993). This correlation has become the primary means for standardizing the luminosity of SNe-Ia. The nature of this correlation has, however, never been sufficiently explained on theoretical grounds. A SN-Ia light curve consisting of pure $^{56}$Ni should exhibit an ``anti-Phillips" relationship where narrow light curves (i.e. those with shorter diffusion times) lose a smaller percentage of their internal energy to adiabatic expansion and thus have higher luminosities (e.g. Woosley et al. 2007 and references therein).

\begin{figure}
\resizebox{\hsize}{!}{\includegraphics{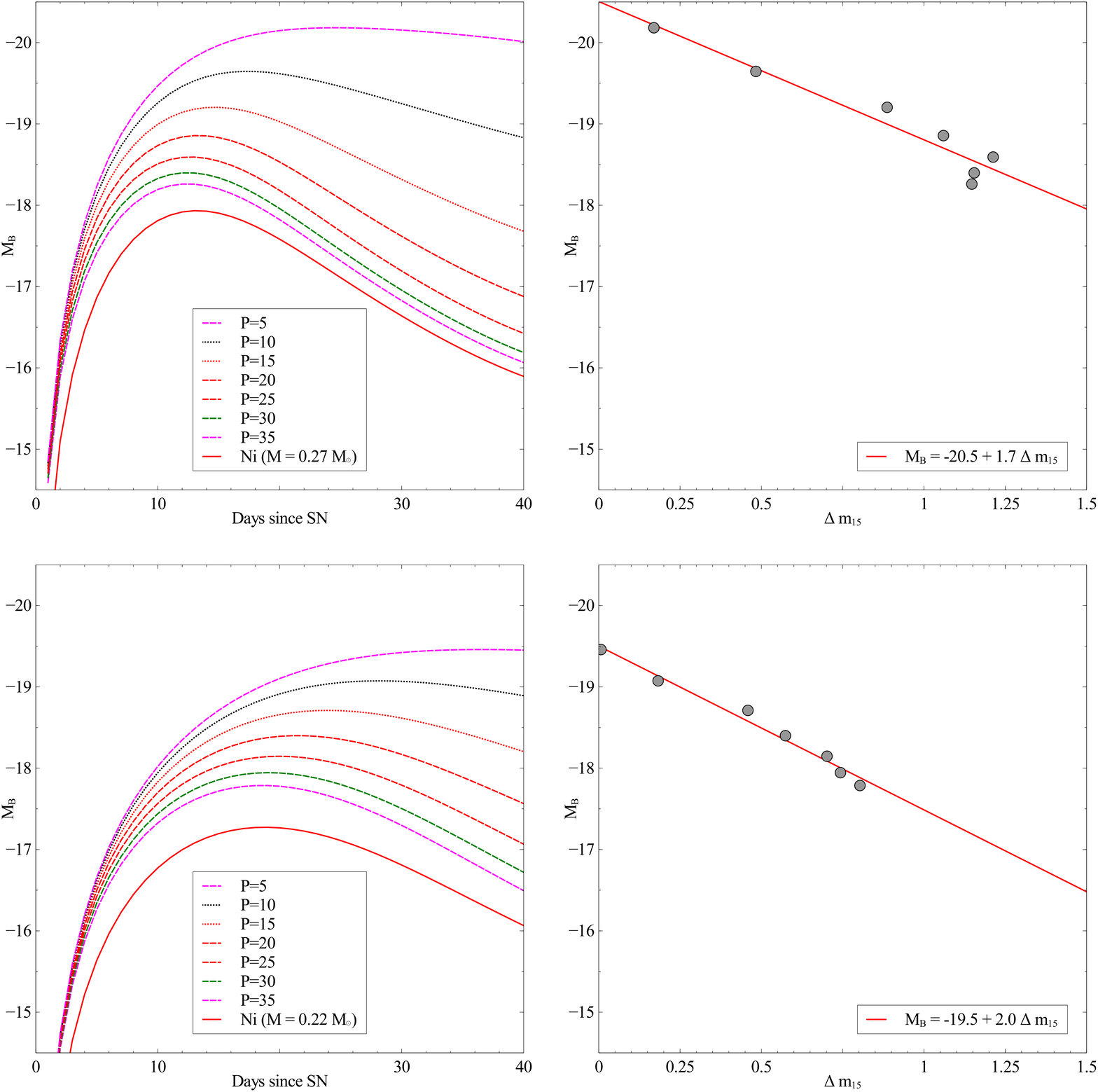}}
\caption{\textbf{Effect on the QN-Ia light curve with varying QS spin-period.}  Examples of QN-Ia light curves generated for various periods using $M_{\rm Ni} = 0.27 M_\odot$ (corresponding to $z = 0$, see \S\ \ref{sec:redshift}) and $M_{\rm Ni} = 0.22 M_\odot$ (corresponding to $z = 1.5$) for the top and bottom panels respectively.  This figure shows qualitatively (left panels) and quantitatively (right panels) that there is a strong correlation between light curve shape and peak absolute magnitude caused by the inclusion of spin-down luminosity.  Furthermore, the form of this correlation is redshift dependent.}
\label{fig:phillips_1}
\end{figure}

The curves in Figure  \ref{fig:phillips_1} qualitatively exhibit a correlation between shape and peak absolute magnitude (left panels of Figure \ref{fig:phillips_1}) and remarkably the correlation between $\Delta m_{\rm B,15}$ and $M_{\rm B,max}$ is almost perfectly linear (right panels of Figure \ref{fig:phillips_1}).  Interestingly, the exact correlation depends on the mass of $^{56}$Ni.
In \S\ \ref{sec:redshift}, we derive a relationship between $M_{\rm Ni}$ and redshift such that $M_{\rm Ni} = 0.27 M_\odot$ corresponds to $z = 0$ and $M_{\rm Ni} = 0.22 M_\odot$ corresponds to $z = 1.5$.  \textit{Figure \ref{fig:phillips_1} therefore shows that the correlation between peak absolute magnitude and light curve shape is redshift dependent.}

\subsubsection{Why $\Delta m_{\rm B, 15}$?} 

In our model, the spin-down luminosity is given as (Deutsch 1955; Manchester \& Taylor 1977):

\begin{equation}
L_{\rm SD} = \frac{E_{\rm SD}}{\tau_{\rm SD}} \left(1+\frac{t}{\tau_{\rm SD}}\right)^{-2} \ ,
\label{eq:LSD}
\end{equation}

where the spin-down energy is:

\begin{equation}
E_{\rm SD} \approx 2 \times 10^{50} ~{\rm \erg} ~ P_{\rm{QS}, 10}^{-2} \ ,
\label{eq:ESD}
\end{equation}
 
and the spin-down time-scale is:
 
\begin{equation}
\tau_{\rm SD} \approx 4.75 ~ {\rm \ days} ~ P_{\rm{QS}, 10}^{2} B_{15}^{-2} \ .
\label{eq:tauSD}
\end{equation}

In the equations above the QS period, $P_{\rm QS, 10}$, is in units of 10 ms and the quark star's
magnetic field\footnote{Amplification of the NS magnetic field up to $B_{\rm QS}\sim 10^{15}$ G, or greater, can be achieved as a result of color-ferromagnetism  during the phase transition from Hadronic to quark matter (e.g. Iwazaki 2005; see also Ouyed et al. 2007a\&b). }, $B_{\rm QS, 15}$, in units of $10^{15}$ G.

The luminosity (in units of $10^{43} \rm{~erg~s^{-1}}$) from $^{56}$Ni + $^{56}$Co decay is:

\begin{equation}
L_{\rm Ni} = \frac{M_{\rm Ni}}{M_\odot} \left(6.45 e^{-t_{\rm days}/8.8} + 1.45 e^{-t_{\rm days}/111.3}\right) \ .
\label{eq:LNi}
\end{equation}

\begin{figure}
\resizebox{\hsize}{!}{\includegraphics{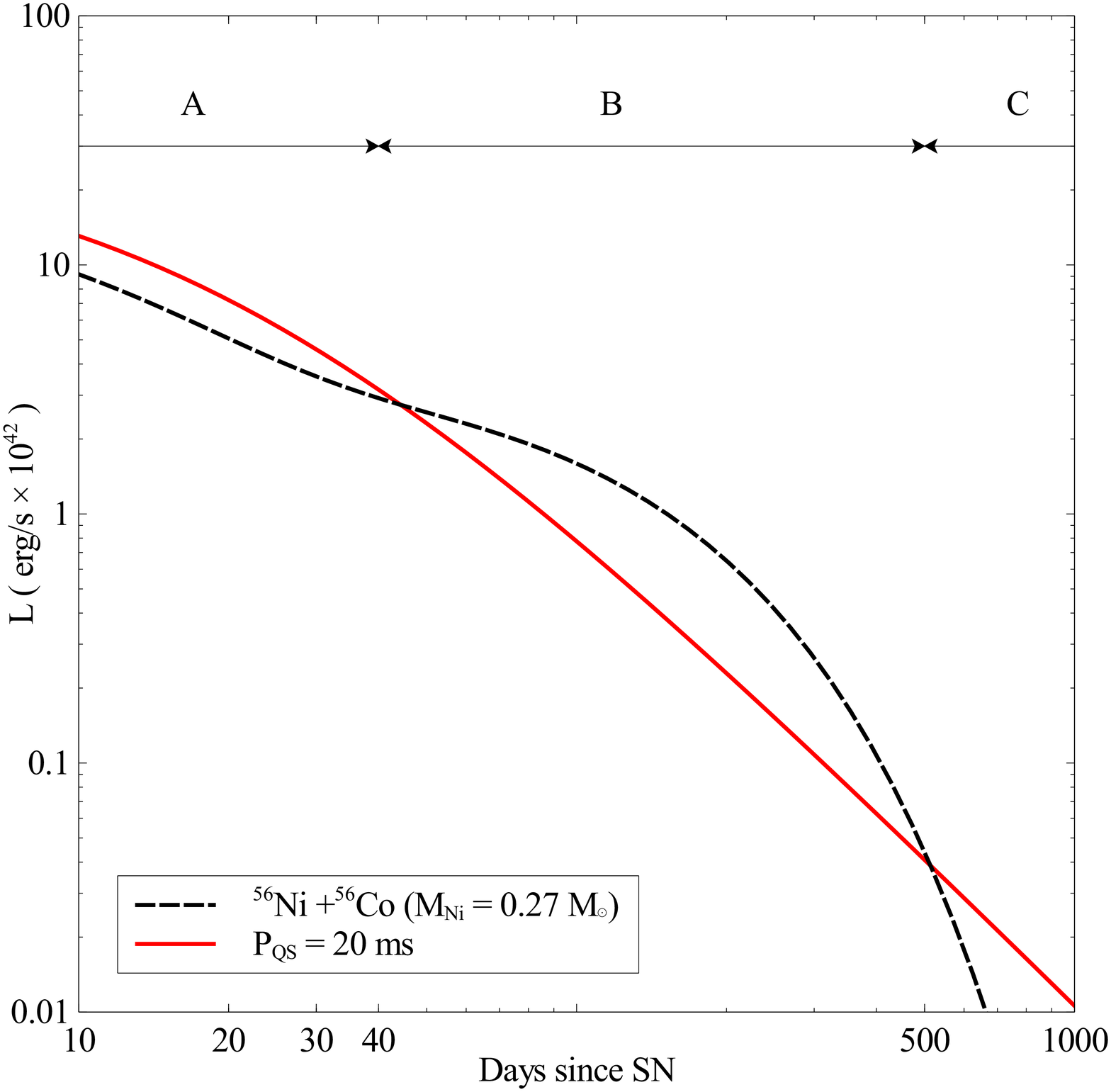}}
\caption{\textbf{Comparison between spin-down and $^{56}$Ni decay luminosities.}  Plotted are the spin-down luminosity using Equation \ref{eq:LSD} ($P_{\rm QS} = 20$ ms) and the $^{56}$Ni + $^{56}$Co decay luminosity using Equation \ref{eq:LNi} ($M_{\rm Ni} = .27M_\odot$) versus days since the SN explosion.  There are three regions in this plot; A ($t < \sim 40$ days) is dominated by spin-down luminosity, B ($\sim 40  < t < \sim 500$ days) is dominated by $^{56}$Ni + $^{56}$Co decay and C ($t > \sim 500$ days) is again dominated by spin-down.}
\label{fig:phillips_2}
\end{figure}

In Figure \ref{fig:phillips_2}, we plot Equation \ref{eq:LSD} for a spin-down of $P_{\rm QS} = 20$ ms and Equation \ref{eq:LNi} for a mass of $0.27 M_\odot$.  There are three distinct regions in this Figure (for different spin-periods, these regions will vary slightly):

\begin{enumerate}[A.]
\item $t < \sim 40$ days where spin-down dominates.
\item $t \sim 40-500$ days where ($^{56}$Ni + $^{56}$Co) decay dominates.
\item $t > \sim 500$ days where spin-down again dominates.
\end{enumerate}

In region A (when $L_{\rm SD} >  L_{\rm Ni}$) we expect to see a Phillips-like correlation between the breadth of the light curve and the peak absolute magnitude.  For a spin-period of 20 ms, this correlation is valid only for the first 40 days after the explosion, or roughly 25 days after the peak (assuming the peak is $\sim 15$ days after the explosion).  For higher spin-periods, this crossing time decreases such that a spin-period $\sim 22$ ms gives a valid correlation for $\sim 15$ days after peak.  Of course we have no a priori reason to assume that all QS periods are $\sim 22$ ms, and not all light curves will peak at 15 days after explosion.  However, perhaps on average, the majority of QN-Ia have these properties.

The bend in the light curve of observed SNe-Ia at $\sim20-30$ days after explosion (where the rapid drop from peak brightness slows down into an exponential decline of the light curve, e.g. Pskovskii 1984) may also be related to the cross-over from region A to B (spin-down dominated to ($^{56}$Ni + $^{56}$Co) decay dominated).  The secondary maximum observed in the R and I bands may be associated with this transition point as well.  This model predicts that a second bend in the light curve should be observed $\sim 500$ days after the peak (transition from region B to C).  Although most SNe-Ia are not observed as late as 500 days past peak, there is some evidence that SN 1991T may indeed show a bend in its light curve $~\sim 600$ days after peak (Schmidt et al. 1994).


\section{Red-shift Relationships}
\label{sec:redshift}

The shape and peak of the QNe-Ia light curves depend heavily on the QS spin-period, the mass of the WD and the amount of $^{56}$Ni produced in the explosion.  It is important for cosmology, then, to understand how these parameters change with red-shift. Of course a range of values for these parameters will exist at any given red-shift, but the peak of the distribution should be well defined.

\subsection{Progenitor mass vs red-shift relation}
If we assume that the majority of star formation occurred during a period $\sim z_{\rm burst}$, then we can determine the mass of the WD progenitor involved in the QN-Ia at any given $z$. Solving Equation \ref{eq:tms} and using $\tau_{\rm M_2} = \tau(z) - \tau(z_{\rm burst})$ with Equation \ref{eq:tz} we have:

\begin{equation}
M_2 (z) \sim \left(\frac{1}{(1+z)^{3/2}} - \frac{1}{(1+z_{\rm burst})^{3/2}}\right)^{-1/\alpha} M_\odot \ .
\label{eq:M2z}
\end{equation}
Hereafter we, again, consider the average value of  $\alpha =3.5$. 

\subsection{WD mass vs red-shift relation}

The mass of the WD can be found using Equations \ref{eq:MWD_MS} and \ref{eq:M2z}:

\begin{equation}
\frac{M_{\rm WD}(z)}{M_{\odot}} = 0.4 + 0.1 \left(\frac{1}{(1+z)^{3/2}} - \frac{1}{(1+z_{\rm burst})^{3/2}}\right)^{-2/7}  \ .
\label{eq:MWD_z}
\end{equation}

\begin{figure*}
\resizebox{\hsize}{!}{\includegraphics{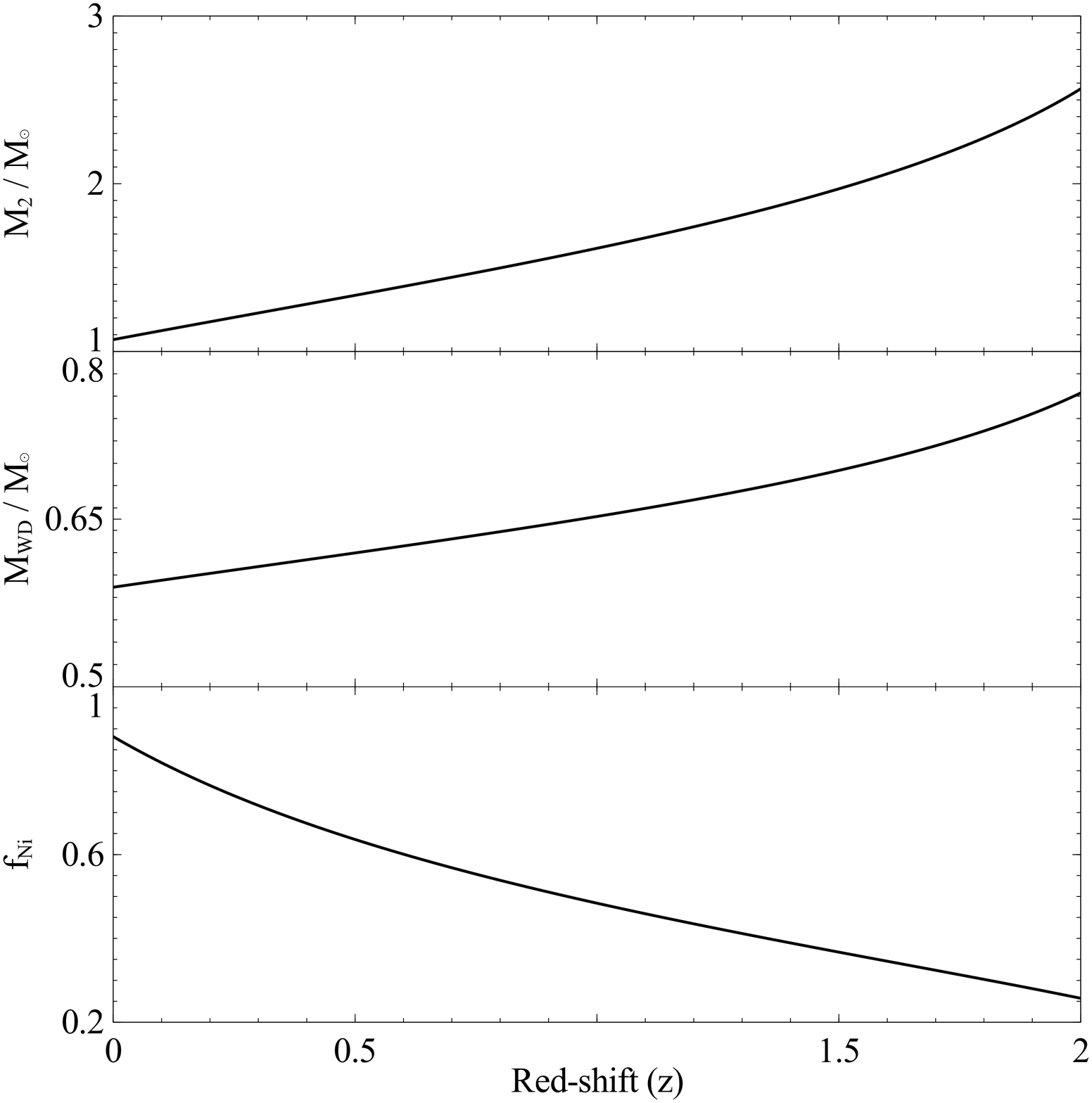}}
\caption{\textbf{Dependence of QN-Ia parameters on red-shift.}}
\label{fig:redshift}
\end{figure*}

Including the effect of metallicity on the WD masses (see Appendix \ref{appendix:Z-Mwd}) gives a dependency on $z_{\rm burst}$ to Equation \ref{eq:MWD_z}:
\begin{eqnarray}
\label{eq:MWDz}
 && \frac{M_{\rm WD}(z)}{M_\odot} \sim e^{0.06 z_{\rm burst}}\times \\\nonumber 
&& \times \left[0.4 + 0.1\left( \frac{1}{(1+z)^{3/2}} - \frac{1}{(1+z_{\rm burst})^{3/2}}\right)^{-2/7}\right] \ .
\end{eqnarray}
The $e^{0.06 z_{\rm burst}}$ factor takes into account the fact that 
stars of lower metallicity  generally produce more massive WDs (see Appendix A).
A plot of Equation \ref{eq:MWDz} is shown in Figure \ref{fig:redshift}.
The starburst lasts for about $\sim 2$ Gyr. This means that nearby QNe-Ia from low-mass progenitors will be nearly uniform in WD mass because the 2 Gy spread is much less than the main-sequence lifetime. On the other hand, high-$z$ QNe-Ia (within 2 Gyr of the starburst) will show a large dispersion in WD mass.

\subsection{$^{56}$Ni mass vs red-shift relation}
\label{sec:Nimass}

We define $T_{\rm C,crit}$ as the critical carbon burning temperature (below which burning to $^{56}$Ni shuts-off) which is reached at time $t_{\rm crit}$ during the expansion \footnote{At $\rho_{\rm WD}\sim 10^6$ g/cc, the Carbon ignition temperature is $T_{\rm C, crit.} \sim 6\times 10^8$ K (Nomoto \& Iben 1985).}.  For an adiabatic expansion, this occurs at:

\begin{equation}
T_{\rm C,crit} = T_0 \left( \frac{R_0}{v_{\rm exp}t_{\rm crit}}\right)^2 \ ,
\label{eq:tcrit}
\end{equation}

\noindent where $T_0$ and $R_0$ ($=R_{\rm WD}$) are the initial temperature and radius of the ejecta\footnote{Ejecta in this context means the post-QN-Ia material (i.e. $M_{\rm WD}-\Delta M_{\rm WD}$), not just the QN ejecta.}  respectively, and $v_{\rm exp}$ is the expansion velocity.  Using Equations 5 and 10 in Ouyed \& Staff (2013) we can express the initial ejecta  temperature as:

\begin{equation}
T_0 \propto \frac{M_{\rm WD, 0.6}^{2/3}}{(0.6 + \zeta^{-2/3})^2 M_{\rm ej,0.4}^{5/3}} \ ,
\label{eq:t0}
\end{equation}

\noindent where $\zeta = M_{\rm WD,0.6}/M_{\rm NS, 1.5}$ with the NS mass given in units of $1.5M_{\odot}$ and $M_{\rm ej,0.4}$ is the mass of the ejecta in units of 0.4 $M_\odot$.  The mass of the ejecta is related to the initial mass of the WD as:

\begin{equation}
M_{\rm ej} = M_{\rm WD} - \Delta M_{\rm WD} \ ,
\label{eq:Mej}
\end{equation}
with
\begin{equation}
\label{eq:dmqn}
\Delta M_{\rm WD} = \Delta M_{\rm QN}-\Delta M_{\rm NS, CE} = \Delta M_{\rm QN}- \psi_{\rm CE} M_2(z) \ .
\end{equation}

In the equation above,  $\Delta M_{\rm WD}$  is the amount of baryonic mass accreted from the WD onto the NS to cause the QN. {\it Recall that $\Delta M_{\rm WD}$ is less than $\Delta M_{\rm QN}$ because of the mass accreted by the NS during the
CE phase ($\Delta M_{\rm NS, CE}$; see \S\ \ref{sec:CE})}.

 The expansion is driven by the energy released during burning (e.g. Arnett 1999):

\begin{equation}
v_{\rm exp}^2 = \frac{2 q M_{\rm Ni}}{M_{\rm ej}} \ ,
\label{eq:vexp}
\end{equation}

\noindent where $q \simeq 0.814~{\rm MeV} / m_H$ is the energy release per nucleon ($m_{\rm H}$ is the hydrogen mass) from carbon burning into $^{56}$Ni (e.g. Arnett 1982).   In our model, $qM_{\rm Ni}/m_{\rm H} >> E_{\rm sd}$ is always true for $P_{\rm QS}> 5$ ms.

Substituting \ref{eq:vexp} and \ref{eq:t0} into \ref{eq:tcrit} and using the WD mass-radius relationship (see e.g. equation 2 in Ouyed \& Staff 2013) we have:

\begin{equation}
\left[\frac{M_{\rm WD, 0.6}^{2/3}}{(0.6 + \zeta^{-2/3})^2 M_{\rm ej,0.4}^{5/3}}\right] \left[\frac{M_{\rm ej,0.4}^{-2/3}}{f_{\rm Ni}^3 M_{\rm ej}^2}\right] = C \ ,
\end{equation}

\noindent where $C$ is a proportionality constant.   {\it To arrive at the equation above we made the assumption
that the amount of $^{56}$Ni processed is proportional to confinement time; i.e. $M_{\rm Ni}\propto t_{\rm crit.}$
 which translates to $t_{\rm crit}\propto f_{\rm Ni} M_{\rm ej}$ with $f_{\rm Ni} = M_{\rm Ni} / M_{\rm ej}$ being the $^{56}$Ni mass fraction}.

Solving for $f_{\rm Ni}$ gives:

\begin{equation}
f_{\rm Ni} = C\frac{M_{\rm WD, 0.6}^{2/9}}{M_{\rm ej, 0.4}^{13/9}(0.6 + \zeta^{-2/3})^{2/3}}\ .
\label{eq:fniz}
\end{equation}

Here  $M_{\rm ej,0.4} $ is the ejecta mass (i.e. $M_{\rm WD}-\Delta M_{\rm WD}$) in units of $0.4M_{\odot}$.
We place an upper limit on $f_{\rm Ni}$ of 1.

Finally using Equations \ref{eq:Mej} and \ref{eq:MWDz} in the equation above gives  the fraction of $^{56}$Ni burnt as a function of red-shift:

\begin{eqnarray}
\label{eq:fniz2}
&& f_{\rm Ni}(z) \simeq  \frac{0.27 C e^{-0.073z_{\rm burst}}}{(0.6 + \zeta^{-2/3})^{2/3}} \times   \\\nonumber
&\times&   \biggl[ 0.4+   0.1\left(\frac{1}{(1+z)^{3/2}} - \frac{1}{(1+z_{\rm burst})^{3/2}}\right)^{-2/7}   -\frac{\Delta M_{\rm WD}}{1\ M_{\odot}}\biggr]^{-13/9}  \\\nonumber
 &\times& \left[0.4 + 0.1\left(\frac{1}{(1+z)^{3/2}} - \frac{1}{(1+z_{\rm burst})^{3/2}}\right)^{-2/7}\right]^{2/9} \ .
\end{eqnarray}

A plot of Equation \ref{eq:fniz2} is shown in Figure \ref{fig:redshift}.

\subsection{QS spin-period vs red-shift relation}

The spin-up of the NS occurs during accretion from the WD in the days preceding the QN.  For $M_{\rm WD} >\sim 0.5 M_\odot$ the mass transfer following RLO is dynamically unstable (e.g. van den Heuvel \& Bonsema 1984; Bonsema \& van den Heuvel 1985; Ritter 1988; Verbunt \& Rappaport 1988).  
The time-scale for the dynamical mass transfer instability as a function of the WD mass was derived in Verbunt \& Rappaport (1988; see \S\ IV.a in that paper).  A fit to their Figure 2 for $0.5M_\odot < M_{\rm WD} < 1.0 M_\odot$ gives the time-scale as:

\begin{equation}
\Delta t_{\rm g} \sim 1\ {\rm week} \times M^{-8.1}_{\rm WD, 0.6} \ ,
\end{equation}

\noindent where the mass of the WD is in units of $0.6M_\odot$.  The corresponding accretion rate is:

\begin{equation}
\dot{m}_{\rm g} \sim \frac{0.6 M_\odot}{1 ~ {\rm week}} M^{9.1}_{\rm WD,0.6} \ .
\label{eq:mdotg}
\end{equation}

 If we assume steady state accretion via  a disk\footnote{Another possibility is that the  NS$+$WD system would enter a CE phase, in which case the NS would spiral  to the core of the WD and explode as a QN.  This path could also lead to a QN-Ia.}, the NS will accrete at a rate  $\dot{m}_{\rm g}$, resulting in hyper-Eddington accretion rates (i.e. $\sim 10^8 \dot{m}_{\rm Edd}$).  Such rates are allowed since it  is in a  regime where neutrino losses dominate (e.g. Chevalier 1993; see also Fryer et al. 1996). 
 
  If the duration of the mass-transfer phase is shorter than the time needed to spin up the NS to spin equilibrium then the NS will not be fully recycled (Gosh \& Lamb 1979). 
  The ratio of torque timescale (e.g. Equation 16 in Tauris et al. 2012) to the accretion timescale $\tau_{\rm acc.}= \Delta M_{\rm WD}/ \dot{m}_{\rm g}$  in our model is 

\begin{equation}
\frac{\tau_{\rm torque}}{\tau_{\rm acc}}\sim 0.6 \frac{\omega_{\rm c, 0.5}}{n_{1.4}} B_{\rm NS, 13}^{-8/7}   \frac{M_{\rm WD,0.6}^{5.2}}{\Delta M_{\rm WD, 0.2}}\ ,
\end{equation}

\noindent where the total accreted mass, $\Delta M_{\rm WD}$ is in units of $0.2M_{\odot}$; the NS radius and mass were set to their fiducial values of 10 km and 1.5$M_{\odot}$, respectively. The critical fastness parameter  $0.25 < \omega_{\rm c}  < 1$   related to the accretion torque (Ghosh \& Lamb 1979) is set to its geometric mean of 0.5 and the dimensionless torque $n$ to 1.4.    
   
The $\tau_{\rm torque}/\tau_{\rm acc}$    spans  a wide range ($<< 1$ to $>> 1$) because of  its dependency
  on NS parameters; the NS magnetic field, $B_{\rm NS}$, being the strongest contributor. Thus for   any given WD mass (i.e. at a given redshift $z$), the NS can have a wide 
range of periods (from fully recycled to partially recycled).  Thus we expect the QS period to vary from a
few milliseconds  to tens of milliseconds. In our model, we use a uniform distribution
of QS periods ranging from $\sim 5$ ms to $\sim 35$ ms. 
 If $P_{\rm QS}$ is too short (i.e. $< \sim 5$ ms), the spin-down energy ($E_{\rm SD}$) will
be absorbed as a {\it PdV} contribution and will affect the rise phase of the LC but  beyond that the
LC will be dominated by $^{56}$Ni decay.  If on the other hand $P_{\rm QS}$ is large ($> \sim 35$ ms),
 then the spin-down contribution is negligible and the LC is again dominated and powered by  $^{56}$Ni decay.
  Thus only QNe-Ia with $5 \ {\rm ms} < P_{\rm QS} < 35$ ms will display a Phillips relationship (i.e.
  will be be accepted by the LC fitters as we show later) while
  those dominated by $^{56}$Ni decay will display  an anti-Phillips behavior (and will be rejected by the LC fitters).

\subsection{Summary}

To summarize this section:

\begin{itemize}

\item  The period distribution, 5 ms $< P_{\rm QS} < $ 35 ms,  of the NS star (and thus of the QS), {\it unlike $M_2(z), M_{\rm WD}(z)$ and $f_{\rm Ni} (z)$, is independent of red-shift}. 

  \item The key dependency on $z$ is the $^{56}$Ni fraction $f_{\rm Ni}(z)$ (eq. \ref{eq:fniz2}). This is controlled by the WD 
  mass dependency on $z$ (eq. \ref{eq:MWDz}) which in turn depends on $M_2(z)$ (eq. \ref{eq:M2z}).
  
  \item The $M_2(z)$ relationship assumes that the gravitational radiation decay timescale, $\tau_{\rm GW}$ (see \S\ \ref{sec:CE})
  is small compared to the main-sequence lifetime of the WD progenitor $M_2$. This is valid for $M_2 < 5 M_{\odot}$.
  From Figure \ref{fig:redshift} (upper panel), this is a good approximation since $M_2 < 5M_{\odot}$ at $z\sim 2$.

\end{itemize}

\section{Hubble Diagram}

Equations \ref{eq:MWDz} and  \ref{eq:fniz2}  allow us to simulate QN-Ia light curves at any red-shift.  Our aim in this section is to compare the actual and fitted distance moduli obtained from these light curves.  Once we have an understanding on how these two distance moduli differ as a function of red-shift, we will examine the implications for the Hubble diagram.

\subsection{The modulus offset: $\Delta \mu (z)$}
\label{sec:offset}

For each simulated QN-Ia, we choose a random $z$ between 0 and 2. Additionally we randomize several other parameters to better simulate the dispersion seen in observations: 

\begin{itemize}
\item $z_{\rm burst}$ between 2.4 and 2.6
\item $M_{\rm NS, b}$ between 1.4$M_{\odot}$ and $1.5M_{\odot}$  {\bf ($M_{\rm NS, c}=1.9M_{\odot}$)}
\item $P_{\rm QS}$ between  5 ms and 35 ms. 
\end{itemize}

Using Equations \ref{eq:MWDz}, \ref{eq:fniz2} (with $C = 0.8$) and \ref{eq:dmqn} (with $\psi_{\rm CE}=0.2$) \footnote{We also remark that  the $\Delta M_{\rm WD}> 0$ condition (see eq. \ref{eq:dmqn}) is satisfied by the   $M_2 < 5M_{\odot}$ for $\psi_{\rm CE}=0.2$.}, we calculate the B and V band light curves as prescribed in \S\ \ref{section:lightcuves}.  Our simulated light curves are all generated in the rest-frame of the QN-Ia, and no stretching of the light curve is performed.  This is so that no K-correction is required by the light curve fitters, thereby reducing the error in the fit.  The red-shift is thus used only to calculate the QN-Ia parameters. We sample the model light curves at 2 day intervals and dim them using a distance modulus of $\mu_0 = 34.035$ (i.e. $z=0.015$).

\begin{figure*}
\resizebox{\hsize}{!}{\includegraphics{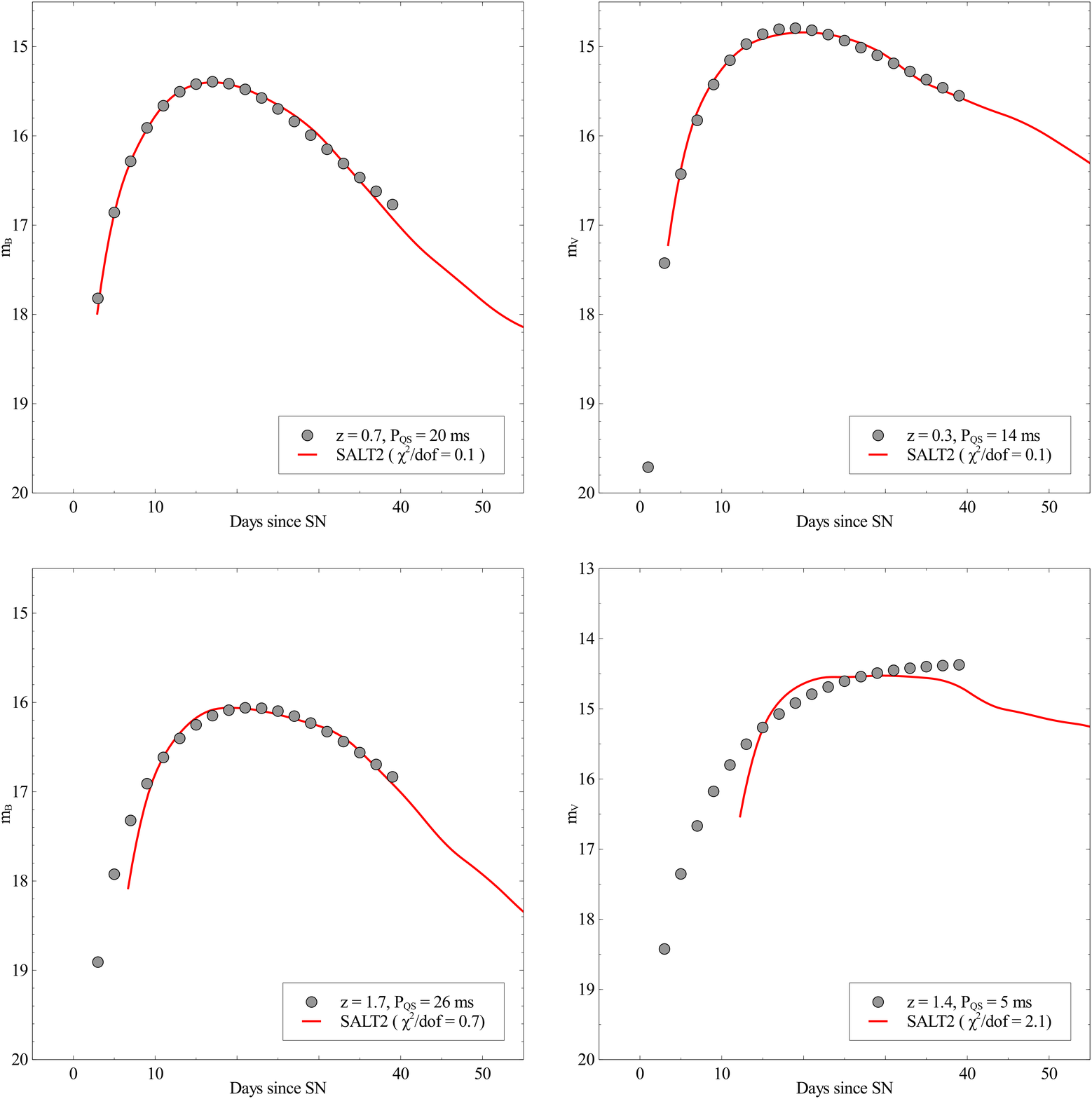}}
\caption{\textbf{SALT2 fits to QN-Ia light curves.}  Shown are simulated QN-Ia light curves for various QS spin-periods and redshifts and the corresponding SALT2 fit.  The $\chi^2/{\rm dof}$ is given for each (we reject light curves with $\chi^2/{\rm dof} > 2$).}
\label{fig:salt2}
\end{figure*}

We use the supernova light curve fitting software SALT2 (Guy et al. 2007) to fit our simulated QN-Ia light curves, examples of which are shown in Figure \ref{fig:salt2}.  For each QN-Ia, SALT2 fits a stretch parameter ($x_1$), the peak B-band apparent magnitude ($m_B$) and a color ($c$) which can then be used to calculate the distance modulus:

\begin{equation}
\mu_{\rm S2} = m_B - M_0 + \alpha x_1 -\beta c \ .
\label{eq:musalt2}
\end{equation}

The parameter $\alpha$ represents the correlation between peak brightness and stretch, $\beta$ the correlation between peak brightness and color and $M_0$ is a magnitude offset.  These parameters are fit by minimizing the residuals in the Hubble diagram using a population of SN-Ia observations.  The values used in this work are those obtained using the Union2.1 compilation of SN-Ia data by The Supernova Cosmology Project (Suzuki et al. 2012): $\alpha$ = 0.121, $\beta$ = 2.47 and $M_0 = -19.321$.  We reject any light curve that is fit with a $\chi^2/{\rm dof} > 2$ or with a stretch parameter $|x_1| > 5$.

\begin{figure*}
\resizebox{\hsize}{!}{\includegraphics{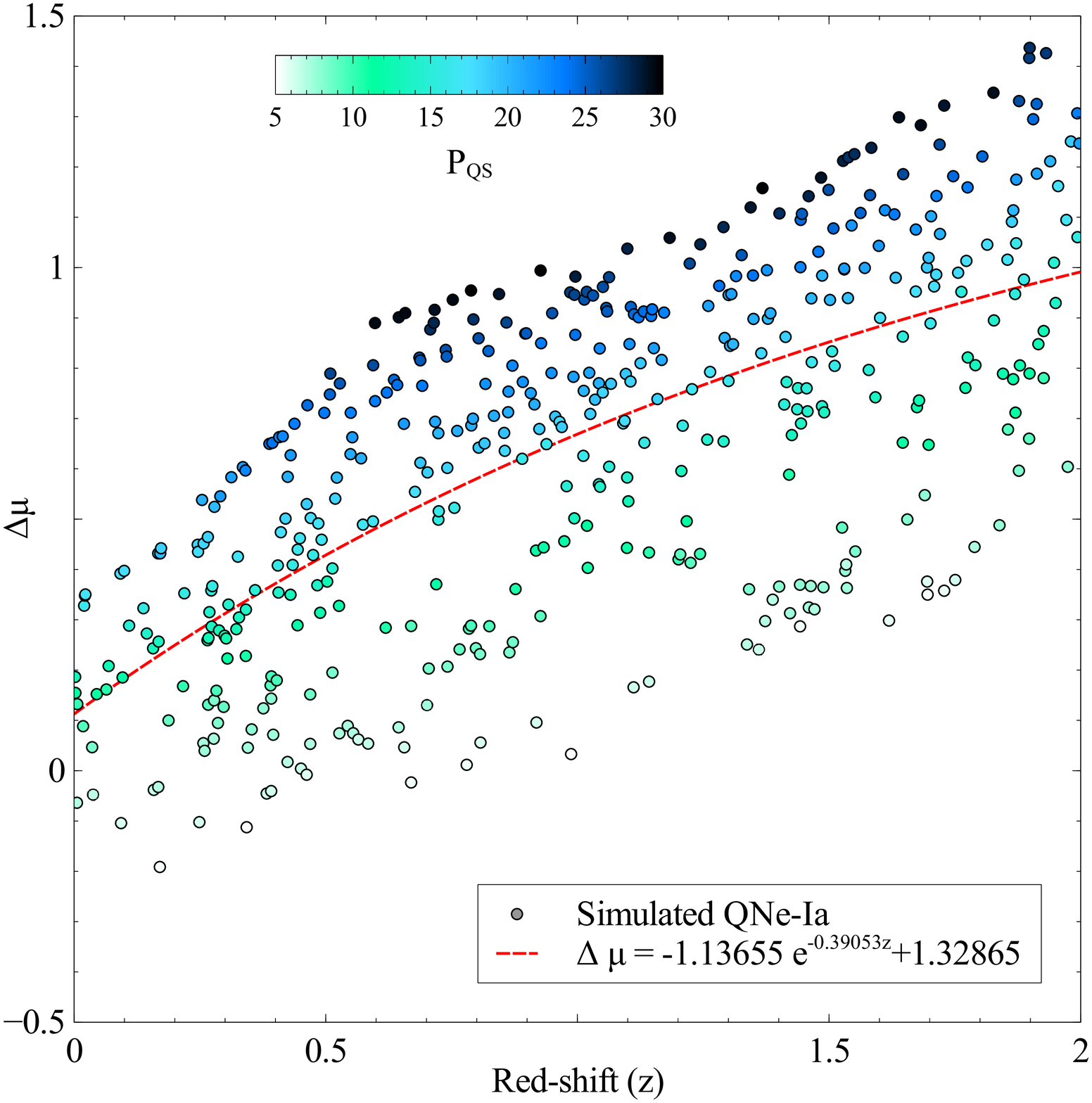}}
\caption{\textbf{Distance modulus correction vs red-shift.}  The difference ($\Delta \mu$) between the true distance modulus and that obtained using the light curve fitters for 500 simulated QN-Ia is plotted for red-shifts between 0 and 2.  This plot demonstrates that the empirical correlations between luminosity and light curve stretch/color are strong at low $z$ but weak at high $z$.  A best-fit curve is shown as a red dashed line and represents the needed ``correction" for observed SNe-Ia.}
\label{fig:dMu_v_z}
\end{figure*}

The difference between the actual distance modulus ($\mu_0$) and the fitted distance modulus ($\mu_{S2}$), $\Delta \mu = \mu_{S2} - \mu_0$, gives us an indication of how accurate the correlations between stretch ($x_1$) and color ($c$) with peak absolute magnitude are at different $z$.  We plot  $\Delta \mu$ vs $z$ for a sample of 500 simulated QNe-Ia in Figure \ref{fig:dMu_v_z}.  The best fit curve to the data is given by:

\begin{equation}
\Delta \mu (z) = -1.13655 \times e^{-0.39053 z} + 1.32865 \ .
\label{eq:deltaMu}
\end{equation}

It is clear that the correlation between stretch, color and peak absolute magnitude is strong at low $z$ ($\Delta \mu \sim 0.19$ at $z = 0$), but is much less accurate at higher $z$ ($\Delta \mu \sim 0.70$ at $z = 1.5$).  This finding has important implications for the Hubble diagram.

\subsection{Hubble Diagram Correction}

\begin{figure*}
\resizebox{\hsize}{!}{\includegraphics{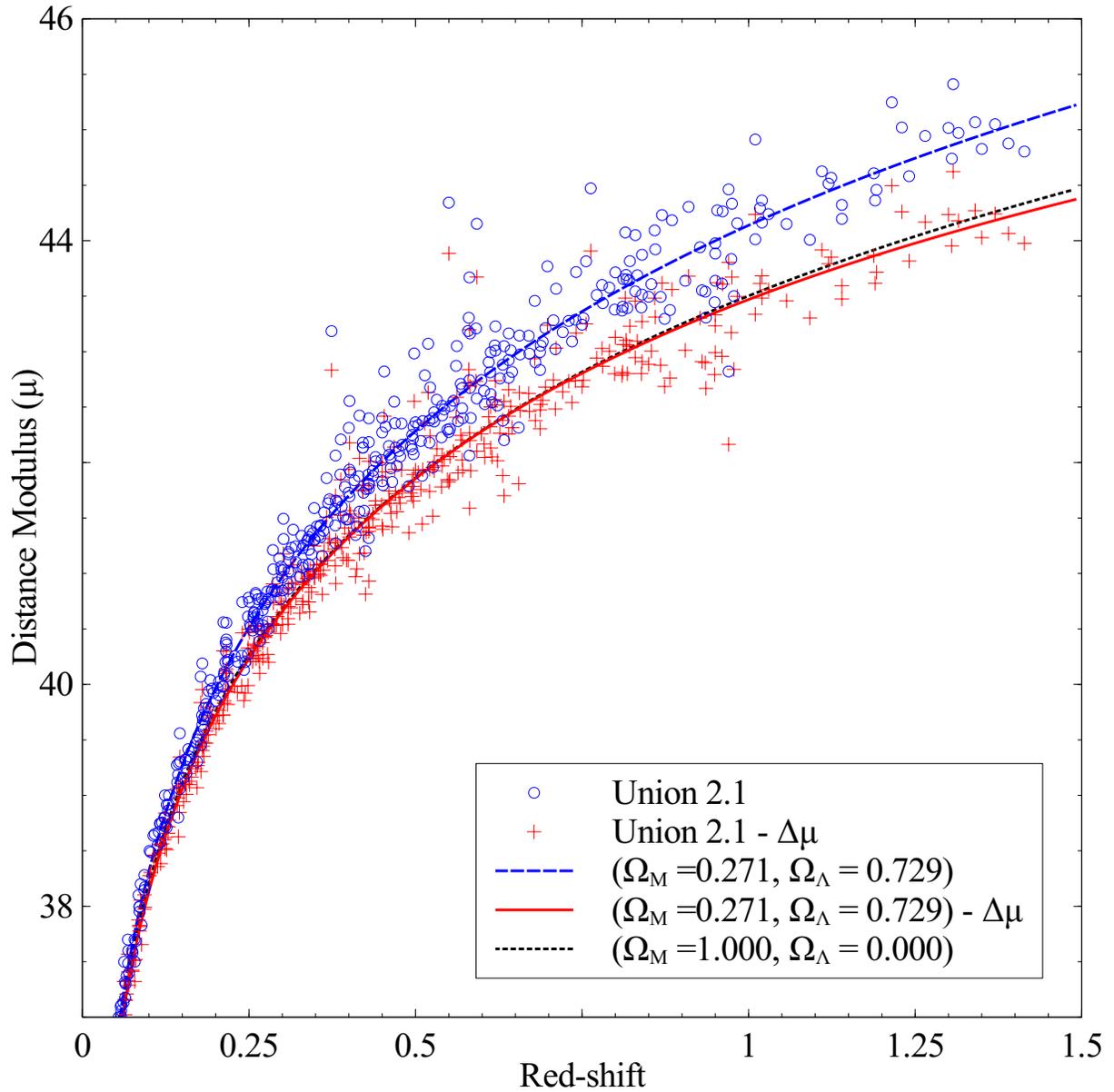}}
\caption{\textbf{QN-Ia Hubble diagram.} Assuming the majority of observed SNe-Ia are in fact QNe-Ia, the distance modulus corrections from Equation \ref{eq:deltaMu} are applied to the Union2.1 data (blue open circles) (Suzuki et al. 2012) resulting in the red crosses.  The corrections are also applied to the best-fit (blue dashed line) cosmology of $(\Omega_{M} =0.271, \Omega_{\Lambda} = 0.729)$ resulting in the red solid curve.  When the QN-Ia corrections are taken into account, the best fit cosmology falls almost exactly along the $(\Omega_{M} =1, \Omega_{\Lambda} = 0)$ curve (black dotted line).}
\label{fig:hubble}
\end{figure*}

The Union2.1 SN-Ia compilation data was fit using SALT2 by Suzuki et al. (2012), and the distance moduli determined using Equation \ref{eq:musalt2} \footnote{Suzuki et al. 2012 include a third weaker correlation between the mass of the host galaxy and the peak absolute magnitude.}.  A plot of $\mu$ vs $z$ for each of the $\sim$ 600 SNe-Ia in the compilation is shown as open blue circles in Figure \ref{fig:hubble}.  The best-fit cosmology of $\Omega_{M} =0.271$, $\Omega_{\Lambda} = 0.729$ (Suzuki et al. 2012) is shown as a blue dashed line.

Applying our correction, Equation \ref{eq:deltaMu}, to the Union2.1 data we obtain the true distance moduli of the observed SNe-Ia (if they are indeed QNe-Ia) as shown by red crosses in Figure \ref{fig:hubble}.  The red-solid line in Figure \ref{fig:hubble} is the result of applying our correction to the best-fit cosmology of $\Omega_{M} =0.271$, $\Omega_{\Lambda} = 0.729$.  For reference we display the $\Omega_{M} =1$, $\Omega_{\Lambda} = 0$ as a dashed black line.  This Figure demonstrates that the SNe-Ia distance moduli, when corrected, lie very close to the $\Omega_{M} =1$, $\Omega_{\Lambda} = 0$ curve.

\section{Discussion}

The stretch (Phillips 1993) and color (Riess et al. 1996; Tripp 1997) of SNe-Ia light curves have been shown to correlate with their peak absolute magnitude.  These empirical corrections allow for the standardization of SNe-Ia and ultimately led to the discovery that the expansion of the Universe is accelerating.  The inherent assumption is that these correlations do not change with time (i.e. red-shift).  Indeed the current theoretical model explaining SN-Ia explosions is not expected to vary with red-shift.  However, our understanding of how these Chandrasekhar (and sub-Chandrasekhar) explosions occur is far from complete.  Ouyed \& Staff (2013) proposed a novel solution to the detonation problem by considering a QN in a tight NS-COWD binary system (QN-Ia), with the resulting explosion mimicking both the spectrum and light curve of a typical SN-Ia.  There are two main differences between QNe-Ia and classical SNe-Ia: First, the spin-down of the QS injects extra energy into the explosion and second, as we have shown in this paper, the parameters governing the QN-Ia are not constant in time, but vary with red-shift. This second point contradicts the assumption made when standardizing SN-Ia explosions and, as we have shown here, when taken into consideration,  relieves the need for DE to explain the SN-Ia observations.

This conclusion relies primarily on two hypotheses:
\begin{enumerate}
\item    The first is that a QN is an endpoint  for massive stars (Ouyed et al. 2002).   The  conjecture  that up-down-strange (UDS) matter is more stable than hadronic matter (Witten 1984) led many authors to  consider that the core of a NS may convert to this more stable form of quark matter  (e.g. Weber 2005 and references therein).  Hydrodynamical simulations of this conversion process have shown that a detonation may occur (Niebergal et al. 2010; see Ouyed et al. 2013
for a review) and when coupled with gravitational collapse result in a QN.  Additionally, much work has been done investigating the observable consequences of QNe.  Phenomena such as SGRs, AXPs, Gamma-Ray Bursts (GRBs), super-luminous SN and double-humped SNe can all be well explained by considering the QN hypothesis.  Furthermore QNe provide an efficient site for the r-process (Jaikumar et al. 2007) and a solution to the cosmic lithium problem (Ouyed 2013b).  Although this does not prove the existence of QNe, it does provide sufficient motivation for their consideration.

\item    The second hypothesis made in the work is that QNe-Ia represent the majority of SNe-Ia explosions.  If only a small fraction of SNe-Ia were in fact QNe-Ia, then the Hubble diagram would not be sufficiently affected to warrant our conclusions. The expected rates of QNe-Ia,  may not be too different from the
observed rates of SNe-Ia (see \S\ \ref{sec:rates} and a discussion in \S\ \ref{sec:limitations}). This  suggests that there may be   enough QNe-Ia to fill the Hubble diagram.

 \end{enumerate}

 There are challenges with the accepted  SN-Ia explosion scenarios such as 
   the source and cause of ignition which is not yet known (e.g. R\"opke et al. 2012).  
The QN-Ia model presented in Ouyed \& Staff (2013) and extended in this work can provide a cure.  First, the detonation of the WD in the QN-Ia scenario is simply explained by shock physics governing the interaction of the QN ejecta and the WD; i.e.  the QN-Ia provides a direct route 
   to ignition.  Secondly, the QN-Ia provides an elegant explanation for the correlation between peak magnitude and light curve shape through the addition of spin-down energy from the QS. Furthermore, a QN-Ia is still an explosion of a WD, and thus is  difficult to distinguish from a thermo-nuclear Chandrasekhar and/or sub-Chandrasekhar explosion.

\subsection{Outliers in our model}

For $P < 5$ ms the spin-down energy  exceeds the C-burning energy ($q M_{\rm Ni}/m_{\rm H}$)
  which results in higher ejecta expansion velocity. Furthermore, a short period 
  means much broader B-band LCs.   These properties can account
   for SNe-Ia outliers like SN 1991T  (see \S\ 4.3 in Ouyed \& Staff (2013)).

 \subsection{The Baryon Acoustic Oscillations}
 
  Modern SNe-Ia surveys  treat systematic errors very thoroughly and
 conclude that a large DE contribution to Cosmology is required  (e.g. Kowalski et al. 2008; Amanullah 2010; Suzuki et al. 2012 ). 
 As an independent  cross-check of this result 
 the most recent  mapping of the distance-redshift relation with baryon acoustic oscillations (the WiggleZ Dark Energy Survey ;
Blake et al. 2011)  are consistent with a cosmological constant model. 
 This appears to disagree with our model and our finding that $\Omega_{\rm M}=1$ is allowed.
  This issue is beyond the scope of this paper, but is important and will be addressed in future work.
 For now, we simply  point out that  that the existence of QNe-Ia does not contradict the concordance model.
Our simulated light curves simply mean that the current cosmological constraints derived from SNe-Ia
surveys need to be re-evaluated.  Our work questions the standard interpretation
 that  SNe-Ia surveys imply DE. However, DE is still allowed, but not required, under our model.

 \subsection{Model Limitations}
 \label{sec:limitations}

{\it (i)}:     By assuming $M_{\rm NS,CE}\sim 0.2M_2$ we effectively considered that the 
    Bondi-Hoyle accretion rate is above the  limit for spherical neutrino cooled accretion (Chevalier 1993;
    see also Chevalier 1996). However whether neutrino cooled accretion can occur during spiral in through the envelope is
    still a matter of debate (Fryer \& Woosley 1998;  Ricker \& Taam 2012).
    
 {\it (ii)}:   The CE channel  is a  mass reservoir  that helps increase the NS mass   before it accretes  from the WD.
  Too much mass accreted  during this phase means the QN can occur before the WD 
  accretion phase and too little mass accreted during the CE phase  means the NS would need 
   to accrete most of the WD mass, particularly if $M_{\rm NS,c}$ is on the high side. For example, 
   if $M_{\rm NS, c}=2.0 M_{\odot}$ then  a typical $1.4M_{\odot}$ NS would need to  accretes 0.6$M_{\odot}$
of  material to experience a QN explosion. In our prescription, at low-$z$,  the  NS would accrete
 $\sim  0.2M_{\odot}$ (since $M_2\sim 1M_{\odot}$)  during the CE phase,
    and would rely on a smaller WD reservoir ($M_{\rm WD}\sim 0.55M_{\odot}$) to make up for the extra 
     $\sim 0.4M_{\odot}$ needed to go QN. In this scenario,  the low-$z$
  QNe-Ia would be associated with the explosion of   about $0.15M_{\odot}$  of WD material (the ejecta).  
  However, this is not necessarily a limitation since the  compression of the remaining WD material following impact by the QN ejecta  would lead 
   to thermo-nuclear burning   under degenerate conditions (see Ouyed \& Staff 2013). 
   Furthermore, as we have found,  at low-$z$    the low-mass ejecta means  complete burning to $^{56}$Ni (i.e. $f_{\rm Ni}=1$; see \S\ \ref{sec:Nimass}) which leads   to smaller $\Delta \mu$ values (see \S\ \ref{sec:Nimass}).

{\it (iii)}:  The given current SNe-Ia rate is  $10^{-5}\ {\rm Mpc}^{-3}\ {\rm yr}^{-1}$
or $\sim 10^{-3}\ {\rm SNe}\ M_{\odot}^{-1}$ (e.g. Dilday et al. 2008; Maoz \& Mannucci 2012; Quimby et al. 2012).
 It translates to $\sim 10^{-3}\ {\rm SNe}$ per year per galaxy when assuming 
a star   formation rate of a few $M_{\odot}$ per year per galaxy.   The QN-Ia rate can be made reasonably close to the observed SNe-Ia rate,
for example by increasing the number of binaries surviving the SN explosion and by
 increasing the number of in-spirals which experience a QN-Ia phase (see eq. \ref{eq:fQN-rate}).
  However these numbers remain to be confirmed before reaching a firm conclusion.

  {\it (iv)}:  The dependence of the $^{56}$Ni fraction on redshift (eq. \ref{eq:fniz2}) was derived by  assuming that the amount
   of $^{56}$Ni produced during the explosion  is proportional to confinement time (see \S\ \ref{sec:Nimass}).  
    It is necessary to perform detailed multi-dimensional  simulations of the explosion (i.e. coupling the hydrodynamics with a  nuclear
     network code) in the context of a QN-Ia  to better constrain this dependency.  In our model, we require 
      that   $f_{\rm Ni}$  decreases with redshift.

\section{Conclusion and Predictions}

We have shown that QNe-Ia produce SN-Ia-type explosions (i.e. the thermonuclear explosion of a COWD). In addition to $^{56}$Ni decay,
 a QN-Ia is also powered by spin-down energy. This gives the result that  QNe-Ia obey a Phillips-like relation
 where the variation in luminosity is due to spin-down power. We also find the calibration relation
to be redshift-dependent (primarily because the $^{56}$Ni mass varies with redshift) which means that QNe-Ia are not standard candles.  
 Applying SN-Ia LC fitters  to QNe-Ia gives a peak magnitude which is different than the true magnitude
 which means distance moduli from LC fitters is offset from the intrinsic one  (i.e. an offset of $\Delta\mu$).
  In our model, the $^{56}$Ni mass decreases with redshift whereas the periods retain a wide
  distribution at all redshifts. This means at higher redshifts QNe-Ia are fainter and thus the $\Delta\mu$
   from the LC fitters will be higher at higher redshifts.  Correcting for this shift in distance modulus  in the Hubble diagram, 
  we find QNe-Ia to be consistent with an $\Omega_{\rm M}=1$ flat universe.  The implication
  is that if the observed SNe-Ia are dominated by QNe-Ia then there might not be a need for DE.
   Our estimated QNe-Ia rate is not too different from  the observed total rate of SNe-Ia which makes this   a real possibility.

\subsection{Predictions}
\label{sec:predictions}

A list of predictions have been given in \S\ 7 in Ouyed \& Staff (2013). Here we reiterate the key ones  and 
  list  new ones:
  
  {\it (i)}:   The  spun-up NS, in the days preceding  the QN-Ia,  might be seen as a radio pulsar.
  
  {\it (ii)}: The scenario predicts a compact remnant (the QS) near the center of Type-Ia remnants with possibly an offset
 determined by a kick.  The QS is an aligned rotator (Ouyed et al. 2004; Ouyed et al. 2006) and will not be seen as a radio pulsar. However, 
  it will emit  X-rays      following vortex (and thus magnetic field) expulsion  and its decay via magnetic reconnection (see eq. 2 in Ouyed et al. 2010 and references therein).   The X-ray luminosity from spin-down may be below the detection limit in historical SNe-Ia.
    For the  Tycho's SN-Ia   with an age of $\sim 441$ years, and assuming an average  birth period of $15$ ms
   for the QS, we estimate a spin-down luminosity  of $\sim  10^{35}$ erg s$^{-1}$; this is an upper value since the QS
    magnetic field would have decayed in time  because of vortex expulsion (see eq. 2 in Ouyed et al. 2010 and references therein)
     and its current period higher than its birth period.
     These limits would also apply to Kepler's supernova since these two historical SNe-Ia are roughly of the same age.
   There is also the possibility of the QS turning into a Black hole   in cases  where enough mass
     from the WD  manages to fall onto the QS.

  {\it (iii)}: The  hyper-accretion   rate,  directly leading to the QN explosion,  should generate temperatures high enough for strong neutrino emission. 
 A  luminosity  of  the  order $0.1M\odot /1\ {\rm week} \sim 10^{46}$-$10^{48}$ erg $s^{-1}$ in tens of MeV neutrinos would not be unreasonable.
  
  {\it  (iv)}:  Gravitational waves (GW) detectors should see signatures of two explosions, the QN explosion and the WD detonation (a fraction of a second apart). The QN GW signature have been investigated in  Staff et al.  (2012).

{\it  (v)}: QNe-Ia should also occur in NS-HeWD binary systems. The burning of the He WD following
impact by the QN ejecta would lead to a Type Ib/Ic explosion and hint at natural connections to GRBs (Ouyed et al. 2011a; Ouyed et al 2011b).

{\it  (vi)}:  The neutron- and  iron-rich QN ejecta was shown to be an ideal site for the
 nucleo-synthesis of heavy elements, in particular nuclei
 with atomic weight $A>130$  (Jaikumar et al. 2007).   Despite the fact that the
  QN ejecta mass is small (on average $\sim 10^{-3}M_{\odot}$) compared to the CO
  ejecta,  these nuclear proxies should be distinguishable in the late spectrum of SNe-Ia
  if these were QNe-Ia.

\subsection{Other evidence for QNe}

Signatures of QNe from single massive stars were predicted and observed. In particular,
 the predicted  dual-hump LC  
 where the first hump  is from  the SN proper and the second brighter one from the re-energization, a few weeks later, 
of the SN ejecta by  the QN ejecta (the dual-shock QN; e.g. Ouyed et al. 2009) have been recorded. This double-hump LC
 has been observed  in the well-sampled Type IIn SNe
SN 2009ip and SN 2010mc and was naturally fit by the QN model  (see Ouyed et al. 2013 and references therein).
 Nuclear spallation  in dual-shock QNe (from the interaction of the $\sim$ GeV QN neutrons with the 
 SN ejecta) may solve the discrepancy between  $^{44}$Ti and $^{56}$Ni  abundances in some Type II SNe (Ouyed et al. 2011c) and 
  produces a $^7$Li plateau when applied to  very metal-poor stars (Ouyed 2013a\&b).

\subsection{Final thoughts}

 Extra-ordinary claims require extra-ordinary evidence.  QNe-Ia seem to be an extra-ordinary claim
  because it is unfamiliar to the astronomical community.  However the underlying physics 
  is based on standard Quantum-Chromo-Dynamics with a plausible assumption about the
  stability of UDS matter.  The  astrophysics involved is not exotic
  and  utilizes properties of compact stars, gravitational collapse, and explosive nuclear burning. 
   On the other hand, DE is truly extra-ordinary  and requires an entirely
  new physics and a new universe.     We are facing a conundrum: for many neither option  is appealing.     Although BAO supports  DE,  this works shows that one has to use caution  when using SNe-Ia   as standardizable candles for cosmology.

\begin{acknowledgements}   

D.T.C.  thanks R.O. for hospitality during the start of this project. We thank D. L. Welch  for useful discussions. This research is supported by  operating grants from the National Science and Engineering Research Council of Canada (NSERC). N.K. would like to acknowledge support from the Killam Trusts.

\end{acknowledgements}


\begin{appendix}

\section{Effect of Metallicity on WD masses}
\label{appendix:Z-Mwd}

Stars of lower metallicity, $Z_M$, at a given mass generally produce more massive WDs according to stellar evolution calculations (e.g., Umeda et al. 1999).  In decreasing the initial stellar metallicity from $Z=0.03$ to $0.004$, Umeda et al. (1999) determined that on average an additional $\sim10\%$ in mass is added to the CO remnant at fixed initial mass.  A fit to the results in Figure 6 of Umeda et al. (1999) gives:

\begin{equation}
M_{\rm WD} \sim M_{\rm WD, 0.004} \times \left( \frac{\rm Z_M}{0.004}\right)^{-0.1}
\label{eq:MWD_zm}
\end{equation}

\noindent where $M_{\rm WD, 0.004}$ is the WD mass at $Z_M$ = 0.004 which varies with the progenitor's mass $M_{\rm i}$.  Furthermore, fits to the observed shape of the mean-metallicity with red-shift gives an exponential dependence (e.g. Kulkarni \& Fall 2002 and references therein):

\begin{equation}
{\rm Z_M} = 0.23 Z_\odot e^{-0.6z}
\label{eq:zm_z}
\end{equation}

Combining Equations \ref{eq:MWD_zm} and \ref{eq:zm_z} gives us the WD mass dependency on $z$:

\begin{equation}
M_{\rm WD} \sim M_{\rm WD,0}e^{0.06 z_{\rm burst}}
\end{equation}

\noindent where $M_{\rm WD,0}$ is the mass of the WD at $z=0$ ($= 0.67M_{\rm WD, 0.004} $) and set
$z=z_{\rm burst}$ since most of the QNe-Ia progenitors occur at around $z=z_{\rm burst}$.

\end{appendix}

\clearpage

\clearpage


\clearpage

\clearpage

\clearpage

\clearpage

\clearpage

\end{document}